\documentstyle[12pt]{article}
\newcommand{\eq}{\begin{equation}}
\newcommand{\en}{\end{equation}}
\newcommand{\eqn}{\begin{eqnarray}}
\newcommand{\enn}{\end{eqnarray}}

\begin{document}
\begin{titlepage}
\begin{flushright}
hep-th/9306011\\
THEP-93-5 \\
2 June 1993
\end{flushright}
\begin{center}
{\LARGE
Sets and $\bf C^{n}$; Quivers and $A-D-E$; \\
Triality; Generalized Supersymmetry; \\
and $D_{4} - D_{5} - E_{6}$} \\
\vspace{0.5cm}
{Frank D. (Tony) Smith, Jr.} \\
{\small Department of Physics \\
Georgia Institute of Technology \\
Atlanta, Georgia 30332} \\
\vspace{0.2cm}
{\bf Abstract}
\end{center}
The relationship between mathematical Geisteswissenschaft
and physics Naturwissenschaft has been discussed by
M\"{u}nster in hep-th/9305104 \cite{MUN}.

The plan of this paper is to begin with the empty set
$\emptyset$; use it to form sets and quivers (sets of points
plus sets of arrows between pairs of points); and then use them
to make complex vector spaces and to get the $A-D-E$ Coxeter-Dynkin
diagrams.

The $D_{n}$ $Spin(2n)$ Lie algebras have spinor
representations to describe fermions.

$D_{4}$ $Spin(8)$ triality gives automorphisms among
the vector and the two half-spinor representations.

$D_{5}$ $Spin(10)$ contains both $Spin(8)$ and
$E_{6}$ contains both $Spin(10)$ and the two half-spinor
representations of $Spin(10)$, and therefore contains the adjoint
representation of $Spin(8)$ and the complexifications
of the vector and the two half-spinor representations of
$Spin(8)$.
$E_{6}$ is the basis for construction of a fundamental
model of physics that is consistent with experiment
(hep-th/9302030, hep-ph/9301210) \cite{SM}.

\vspace{0.2cm}
\normalsize
\footnoterule
\noindent
{\footnotesize \copyright 1993 Frank D. (Tony) Smith, Jr.,
341 Blanton Road, Atlanta, Georgia 30342 USA \\
P. O. Box for snail-mail:  P. O. Box 430, Cartersville,
Georgia 30120 USA \\
e-mail: gt0109e@prism.gatech.edu
and fsmith@pinet.aip.org}
\end{titlepage}

\newpage

\setcounter{footnote}{0}
\section{Does Naturwissenschaft Physics come from
Geisteswissenschaft Mathematics?}
\setcounter{equation}{0}

In hep-th/9305104, Gernot M\"{u}nster discusses his view of
the role of mathematics in physics.  Since mathematics is
Geisteswissenschaft, a creation of the human mind, and
physics is Naturwissenschaft, the observed phenomena of
the physical world, it is remarkable that mathematics is so
useful in describing physics.  M\"{u}nster is certainly not
alone in his views, as he cites Dyson, Manin, Wigner, Courant,
Hertz, and others who are similarly impressed by the deep
connections between mathematics and physics.

\vspace{12pt}

The purpose of this paper is to try to start with
{\it fundamental} mathematical structures,

then use {\it natural} operations on them to construct more
elaborate structures, and

finally to observe that among these {\it natural fundamental}
structures, there is a unique structure with a triality relation
between spinor fermions and vector spacetime that induces
a relation between spinor fermions and  bivector bosons.

Unlike 1-1 supersymmetry, it does not require superalgebras,
nor does it produce a lot of unobserved particles.

Such a  structure can be used to construct a model of
elementary particle physics and gravity that is consistent
with experiment \cite{SM}.

\vspace{12pt}

The key words above are {\it fundamental} and {\it natural}.
It should go without saying that in this paper I will use
them according to my views.  I certainly recognize that
others have other views and so may disagree with my choices,
and that alternative choices may be as interesting and
useful as mine, or perhaps more so.  However, I think that
the structures set out in this paper show the existence of
at least one reasonable path leading from the mental
constructs of fundamental mathematics to the experimental
results of elementary particle physics plus gravity.

\vspace{12pt}

To show that the answer is YES to the question of this section
"Does Naturwissenschaft Physics come from
Geisteswissenschaft Mathematics?" a path from
Geisteswissenschaft to Naturwissenschaft must be found.

\vspace{12pt}

Since the path must lead to Naturwissenschaft, and since
the minimal $SU(3) \times SU(2) \times U(1)$ standard
model plus gravity has been confirmed by all experiments so
far, the path must be required to lead to a model containing
spacetime, gauge bosons, fermion particles, and fermion
antiparticles, related to each other in a way that is consistent
with the Lagrangian quantum field theory of the minimal
standard model, plus gravity.

\vspace{12pt}

Since the path should begin with Geisteswissenschaft
fundamental mathematics, start with some fundamental
mathematical structure and proceed according the
following outline of the remainder of this paper:

\vspace{12pt}

\begin{tabbing}
2  \= Sets and $\bf C^{n}$ \\
\> 2.1  Sets \\
\> 2.2  Complex n-space $\bf C^{n}$ \\
3  Quivers \\
\> 3.1  Quivers of Sets and Arrows \\
\> 3.2  $\bf C^{n}$ Representation of Quivers \\
4  $A-D-E$ \\
\> 4.1  Ubiquity of $A-D-E$ \\
\> 4.2  $A_{n}$ \\
\> 4.3  $D_{n}$ \\
\> 4.4  $D_{4}$ and Triality \\
\> 4.5  $E_{6}$, $E_{7}$, and $E_{8}$ \\
5  Triality and Generalized Supersymmetry \\
\> 5.1  $A_{n}$ $SU(n+1)$ does not work. \\
\> 5.2  \= What about $D_{n}$ $Spin(2n)$? \\
\> \> 5.2.1  \= Conventional 1-1 supersymmetry, or full spinor - \\
\> \> \> bivector supersymmetry, does not work. \\
\> \> 5.2.2  Half-spinor - bivector supersymmetry does not work. \\
\> \> 5.2.3  Full spinor - vector supersymmetry almost works. \\
\> \> 5.2.4  Half-spinor - vector supersymmetry does work. \\
6  Conclusion:  $D_{4} - D_{5} - E_{6}$
\end{tabbing}

\vspace{12pt}

\section{Sets and $\bf C^{n}$}
\subsection{Sets}

\vspace{12pt}

To start with, take the empty set $\emptyset$ as fundamental.

\vspace{12pt}

The positive integer natural numbers $\bf N$ can be built
from the empty set $\emptyset$ and the set operation $\{  \}$.

\vspace{12pt}

$\bf N$ is built very rapidly in this way.  After the set operation
$\{ \}$ has been used n times, $2 \uparrow 2 \ldots \uparrow 2$
(with n 2s and n-1 uparrows)  numbers have been constructed.

For example, after 5 $\{ \}$ operations have been used,
$2 \uparrow 2 \uparrow 2 \uparrow 2  \uparrow 2$ =

= $2 \uparrow 2 \uparrow16$ = $2^{65,536}$ numbers
have been constructed.

\vspace{12pt}

The integers $\bf Z$, with $0$ and negative integers, then come
from $\bf N$ and the inverse operation of addition, subtraction.

\vspace{12pt}

The rational numbers $\bf Q$ come from $\bf Z$, excluding $0$, and
the inverse operation of multiplication, division.

\vspace{12pt}

The real numbers $\bf R$ come from continuous completion of $\bf Q$.

\vspace{12pt}

The complex numbers $\bf C$ come from algebraic completion
of $\bf R$, and can also be considered as doubling the dimension
of $\bf R$ with the added dimension being imaginary (the basis
element being $\sqrt{-1}$ ).

\vspace{12pt}

The quaternions $\bf H$ come from doubling the dimension of
$\bf C$ with the 2 added dimensions being imaginary,
at the algebraic cost of losing commutativity but preserving
associativity.

\vspace{12pt}

The octonions $\bf O$ come from doubling the dimension of
$\bf H$ with the 4 added dimensions being imaginary,
at the algebraic cost of losing associativity but preserving
alternativity.

\vspace{12pt}

This completes the construction of the division algebras
$\bf R$, $\bf C$, $\bf H$, and $\bf O$ considered to be fundamental
by Geoffrey Dixon \cite{DIX}.

\vspace{12pt}

The complex numbers $\bf C$ are the only topologically complete,
algebraically complete, commutative, and associative division
algebra, so $\bf C$ will be used as the fundamental base field.

\vspace{12pt}

n-dimensional vector spaces $\bf C^{n}$ can be constructed
over $\bf C$.

\vspace{12pt}

Starting with the empty set $\emptyset$, and then proceeding in
a natural way, $\bf C^{n}$ has been built.

\vspace{12pt}

\subsection{Complex n-space $\bf C^{n}$}

\vspace{12pt}

Now that we have $\bf C^{n}$, what do we do with it?
In particular, is there a naturally fundamental way to build
models using vector spaces of the form $\bf C^{n}$?

\vspace{12pt}

Given a bunch of vector spaces
$\bf C^{r}$, $\bf C^{s}$, ... , $\bf C^{t}$,
the most natural thing to do is to define maps among them.

\vspace{12pt}

The $r \times s$ complex matrices $\bf C(r,s)$ describe
$\bf C$-linear transformations from $\bf C^{r}$ to $\bf C^{s}$.

Here, I take $\bf C(r,s)$ as the most natural way to define
maps among the vector spaces of the form $\bf C^{n}$.

I do recognize that some people would say that linear matrices
are too special, and that all sorts of nonlinear maps should be
introduced.

Even so, I will stick with the $r \times s$ complex
matrices $\bf C(r,s)$ and proceed.

\vspace{12pt}

At this stage, I could build $Gl(n,\bf{C})$, and then easily get \\
$SU(3) \times SU(2) \times U(1)$ for the minimal standard model
gauge group; use either $Sl(2,\bf{C})$ or $SU(2,2)$ for the Poincare
group or the conformal Penrose twistor group for gravity; and use
a Hilbert space built from $\bf C^n$ as the space of rays for quantum
state vectors, acted on by complex operators built from complex
Hermitian matrices.

Some might be happy with this, but I am not,
because the mathematical structure does not clearly define
the relations among the gauge groups, the spacetime,
and the quantum states and operators, much less the identities and
masses of fermion particles and antiparticles, the relative strengths
of the forces, and the reason for stopping at $SU(3)$ rather than
including a gauge group such as, say, $SU(17)$.

\vspace{12pt}

\section{Quivers}
\subsection{Quivers of Sets and Arrows}

\vspace{12pt}

To get more structure, I must start with something more than the
empty set $\emptyset$.

First, add an arbitrary set P of points.  Just a set of points by itself
does not add much structure, so include a minimal amount of relations
among the points in the set.

As a minimal set of relations among the points in P,
use a set A of arrows between pairs of the points in P,
with every arrow $\alpha$ in A going from a tail t($\alpha$) in P to
a head h($\alpha$) in P.

Such a 4-tuple (P, A, t, h), where P is a set of points, A is a set of
arrows between points, and, if $\alpha$ is an arrow, t($\alpha$) is
the point at the tail of the arrow and h($\alpha$) is the point at the
head of the arrow, has been studied in 1972 by Peter Gabriel
\cite{GAB}, who thereby founded the theory of quivers of arrows.

\vspace{12pt}

A quiver (P, A, t, h) defines, and is defined by,
both an (unoriented) graph $\Gamma$
(by the points and their connections by arrows) and
an orientation $\Lambda$ (by the directions of the arrows).

Clearly, many quivers may correspond
to the same (unoriented) graph $\Gamma$.

A quiver is called connected if its graph $\Gamma$ is connected.

\vspace{12pt}

\subsection{$\bf C^{n}$ Representation of Quivers}

\vspace{12pt}

Gabriel represented a quiver (P, A, t, h) by representing the
points in P as vector spaces and the arrows $\alpha$ in A as
matrix maps from the vector space representing t($\alpha$)
to the vector space representing h($\alpha$).

\vspace{12pt}

Here, we have the complex vector spaces $\bf C^{n}$ and
the $r \times s$ complex matrices $\bf C(r,s)$ with which to
construct such representations.

\vspace{12pt}

Gabriel's theorem is:

If a connected quiver has only finitely many non-isomorphic
indecomposable representations, its graph is a Coxeter-Dynkin
diagram of one of the Lie algebras $A_{n}$, $D_{n}$, $E_{6}$,
$E_{7}$, or $E_{8}$, and there is a 1-1 correspondence between
the classes of isomorphic indecomposable representations and
the positive roots of that Lie algebra. \cite{GAB}

\vspace{12pt}

\section{$A-D-E$}

Using Gabriel's theorem, we now have the remarkable result
that, if we start with the empty set $\emptyset$ and quivers of
arrows, the only finitely representable structures we have are the
$A-D-E$ Lie algebras.

\vspace{12pt}

\subsection{Ubiquity of $A-D-E$}

\vspace{12pt}

V. I. Arnold \cite{ARN} has pointed out that the $A-D-E$
classification appears in such apparently (but not really)
diverse areas as critical points of functions, Lie algebras,
categories of linear spaces, caustics, wave fronts,
regular polyhedra in 3-dimensional space, and
Coxeter crystallographic reflection groups.

\vspace{12pt}

Robert Gilmore \cite{GIL} has a nice description of how the
$E_{6}$, $E_{7}$, and $E_{8}$ Coxeter-Dynkin diagrams
correspond to the tetrahedron, octahedron, and icosahedron
in 3-dimensional space.  His book is a good introduction
and source-book for Lie groups and related topics.

\vspace{12pt}

Michio Kaku \cite{KAK} describes how the $A-D-E$
classification appears in superstring conformal field theory,
being in 1-1 correspondence not only with the modular
invariants of $SU(2)_{k}$, but also with the special
solutions of solutions of c=1 theory for two continuous
classes and the three discrete solutions.  Kaku says that
this is because of the correspondence between the simply
laced groups and the finite subgroups of $SU(2)$.

Kaku's description indicates that, at this point along the path
from Geisteswissenschaft to Naturwissenschaft, superstring
conformal field theory has a claim to be a natural physics
theory derived from fundamental mathematics.

I do not choose to use superstring conformal field theory as my
example of a fundamental theory for two reasons:

first, there is no natural way I know of to pick a unique
theory from the many possible superstring theories; and

second, all the superstring theories known to me predict
the existence of a lot of particles (many of them arising
from a 1-1 boson-fermion supersymmetry) that have
never been experimentally observed.

\vspace{12pt}

Therefore, after noticing a lot of interesting mathematics
along the way, I will now proceed along my chosen path.

\vspace{12pt}

\subsection{$A_{n}$}

\vspace{12pt}

For $A_{n}$, the $SU(n+1)$ Lie algebra,
the Coxeter-Dynkin diagram has n nodes and n-1 lines:
$${\rm o} \rule[2pt]{0.17in}{0.02in} {\rm o}
\rule[2pt]{0.17in}{0.02in} {\rm o}
\rule[2pt]{0.17in}{0.02in} {\rm o} \cdots
\rule[2pt]{0.17in}{0.02in} {\rm o}$$
Each node corresponds to one of the n fundamental representations
of $A_{n}$ $SU(n+1)$.

All irreducible representations of $A_{n}$ $SU(n+1)$ can be
formed from tensor products and linear combinations of the
n fundamental representations (plus the trivial
1-dimensional "scalar" representation denoted by 1).

If the node on one end (by symmetry, it doesn't matter which end)
of the $A_{n}$ Coxeter-Dynkin diagram corresponds to the
fundamental (n+1)-dimensional representation denoted
by $(n+1)$, then the n-1 other fundamental representations
have dimension given by antisymmetric exterior products of
the (n+1)-dimensional representation space

$(n+1) \wedge (n+1) \ldots \wedge (n+1)$,

and the fundamental representations of $A_{n}$
$SU(n+1)$ can be represented by the non-scalar exterior
products of the complex vector space $\bf C^{n+1}$, which
in turn can be represented by the non-unit terms of the
(n+1) level of the Yang Hui triangle:

\[
\begin{array}{ccccccccccccccccccccc}
&&&&&&&&&&1&&&&&&&&&&\\
&&&&&&&&&1&&1&&&&&&&&&\\
&&&&&&&&1&&{\bf2}&&1&&&&&&&&\\
&&&&&&&1&&{\bf3}&&{\bf3}&&1&&&&&&&\\
&&&&&&1&&{\bf4}&&{\bf6}&&{\bf4}&&1&&&&&&\\
&&&&&1&&{\bf5}&&{\bf10}&&{\bf10}&&{\bf5}&&1&&&&&\\
&&&&1&&{\bf6}&&{\bf15}&&{\bf20}&&{\bf15}&&{\bf6}&&1&&&&\\
&&&1&&{\bf7}&&{\bf21}&&{\bf35}&&{\bf35}&&{\bf21}&&{\bf7}&&1&&&\\
&&1&&{\bf8}&&{\bf28}&&{\bf56}&&{\bf70}&&{\bf56}&&{\bf28}&&{\bf8}&&1&&\\
\end{array}
\]

In the above triangle, the bold-face entries represent the
fundamental representations of the $A_{n}$ $SU(n+1)$ Lie
algebra through n+1 = 8.

The triangular representation of the expansion of $(1+1)^{n}$
was invented by Yang Hui in 1261 A.D., when China was divided
between the Southern Song Dynasty and the Mongol Yuan Dynasty
\cite{MHB}.  It is widely known among Europeans as Pascal's
triangle.

\vspace{12pt}

The adjoint representation of $U(n+1)$ is just the tensor
product \\
$(n+1) \otimes (n+1)$ of the vector and pseudovector
fundamental representations of $A_{n}$, and
$$Adjoint(U(n+1)) = (n+1) \otimes (n+1) =
Adjoint(SU(n+1)) \oplus 1$$

$U(n+1)$ can be represented in $\bf C((n+1),(n+1))$ as the  \\
$(n+1) \times (n+1)$ anti-hermitian complex matrices,
with its $SU(n+1)$ subgroup being represented by those
anti-hermitian complex matrices with determinant 1.

\vspace{12pt}

For example, $A_{7}$ $SU(8)$, with Coxeter-Dynkin diagram
$${\rm o} \rule[2pt]{0.17in}{0.02in} {\rm o}
\rule[2pt]{0.17in}{0.02in} {\rm o}
\rule[2pt]{0.17in}{0.02in} {\rm o}
\rule[2pt]{0.17in}{0.02in} {\rm o}
\rule[2pt]{0.17in}{0.02in} {\rm o}
\rule[2pt]{0.17in}{0.02in} {\rm o}$$

has fundamental representations of dimensions

$$8 \rule[2pt]{0.17in}{0.02in} 28
\rule[2pt]{0.17in}{0.02in} 56
\rule[2pt]{0.17in}{0.02in} 70
\rule[2pt]{0.17in}{0.02in} 56
\rule[2pt]{0.17in}{0.02in} 28
\rule[2pt]{0.17in}{0.02in} 8$$

The 63-dimensional adjoint representation of $SU(8)$ is
determined by

$8 \otimes 8 = 64 = 63 \oplus 1$

\vspace{12pt}

\subsection{$D_{n}$}

\vspace{12pt}

For $D_{n}$, the $Spin(2n)$ Lie algebra, the Coxeter-Dynkin
diagram has n-2 nodes in a row connected with n-3 lines, with
the (by usual convention, right-hand) end node connected
further by 2 more lines to 2 more nodes:
$${\rm o} \rule[2pt]{0.17in}{0.02in} {\rm o}
\rule[2pt]{0.17in}{0.02in} {\rm o}
\rule[2pt]{0.17in}{0.02in} {\rm o} \cdots
\rule[2pt]{0.17in}{0.02in}
\shortstack{o \\ I \\ o}
\rule[2pt]{0.17in}{0.02in} {\rm o}$$

Each node corresponds to one of the n fundamental representations
of $D_{n}$ $Spin(2n)$.

All irreducible representations of $D_{n}$ $Spin(2n)$ can be
formed from tensor products and linear combinations of the
n fundamental representations (plus the trivial
1-dimensional "scalar" representation denoted by 1).

If the node on single node end of the $D_{n}$ Coxeter-Dynkin
diagram corresponds to the fundamental $2n$-dimensional
vector representation of $Spin{2n}$ denoted by $2n$, then
the n-3 other fundamental representations nearest it
have dimension given by antisymmetric exterior products of
the $2n$-dimensional representation space
$2n \wedge 2n \ldots \wedge 2n$
and these n-2 fundamental representations of $D_{n}$
$Spin(2n)$ can be represented by the non-scalar exterior
products of the complex vector space $\bf C^{2n}$, which
in turn can be represented by the first n-2 non-unit terms
of the 2n level of the Yang Hui triangle:

\[
\begin{array}{ccccccccccccccccccccc}
&&&&&&&&&&1&&&&&&&&&&\\
&&&&&&&&&1&&1&&&&&&&&&\\
&&&&&&&&1&&2&&1&&&&&&&&\\
&&&&&&&1&&3&&3&&1&&&&&&&\\
&&&&&&1&&4&&6&&4&&1&&&&&&\\
&&&&&1&&5&&10&&10&&5&&1&&&&&\\
&&&&1&&{\bf6}&&15&&20&&15&&6&&1&&&&\\
&&&1&&7&&21&&35&&35&&21&&7&&1&&&\\
&&1&&{\bf8}&&{\bf28}&&56&&70&&56&&28&&8&&1&&\\
&1&&9&&36&&84&&126&&126&&84&&36&&9&&1&\\
1&&{\bf10}&&{\bf45}&&{\bf120}&&210&&252&&210&&120&&45&&10&&1\\
\end{array}
\]

In the above triangle, the bold-face entries represent the
fundamental vector and multivector (i.e., non-spinor)
representations of the $D_{n}$ $Spin(2n)$ Lie
algebra through 2n = 10.

Since 1-2=-1,  $D_{1}$ $Spin(2)$ has no bold-face entry, as
its 1-dimensional half-spinor representation coincides
with its adjoint representation, and its 2-dimensional
vector representation coincides with its reducible full
spinor representation.

Since 2-2=0, $D_{2}$ $Spin(4)$ has no bold-face entry, as
its 6-dimensional adjoint representation is not
irreducible, $Spin(4)$ = $SU(2) \times SU(2)$, and as
its 4-dimensional vector representation coincides with
its reducible full spinor representation.

Since 3-2=1, $D_{3}$ $Spin(6)$  has no bold-face entry for
the $15$ bivector representation, because it can be
constructed for the two 4-dimensional half-spinor
representations by $15$ = $4$ $\otimes$ $4$  $\oplus$ $1$,
and so is not irreducible with respect to tensor products
and linear combinations.

The adjoint representation of $Spin(2n)$ is just the bivector
exterior product $2n \wedge 2n$, of dimension n(2n-1).

The part of $Spin(2n)$ representable by the n-2 fundamental
representations discussed above can be represented in
$\bf R(2n,2n)$ as the $2n \times 2n$ anti-symmetric
real matrices with determinant 1.

\vspace{12pt}

The two fundamental representations on the far end of
the Coxeter-Dynkin diagram are the half-spinor representations
of $Spin(2n)$, each of dimension $2^{n-1}$.

As each is the mirror image of the other, they are here
denoted by $+2^{n-1}$ and $-2^{n-1}$.

Their sum,  $+2^{n-1}$ $\oplus$ $-2^{n-1}$, has dimension $2^{n}$.
It is the reducible full spinor representation of $Spin(2n)$, and
is here denoted by $2^{n}$.

The square $2^{n}$ $\otimes$ $2^{n}$ of the full spinor
representation of $Spin(2n)$ has dimension $2^{2n}$, and
can be represented by the entire graded Clifford algebra
whose graded dimensions are given by the entire 2n level
of the Yang Hui triangle.

\vspace{12pt}

For example, $D_{5}$ $Spin(10)$, with Coxeter-Dynkin diagram
$${\rm o} \rule[2pt]{0.17in}{0.02in} {\rm o}
\rule[2pt]{0.17in}{0.02in}
\shortstack{o \\ I \\ o}
\rule[2pt]{0.17in}{0.02in} {\rm o}$$

has fundamental representations of dimensions

$$10 \rule[2pt]{0.17in}{0.02in} 45
\rule[2pt]{0.17in}{0.02in}
\shortstack{16 \\ I \\ 120}
\rule[2pt]{0.17in}{0.02in} 16$$

The exterior product of the two half-spinor representations,
$+16$ $\wedge$  $-16$,  has the same dimension, 120, as the
representation to which they are both connected.

\vspace{12pt}

\subsection{$D_{4}$ and Triality}

\vspace{12pt}

The triality relationship among the two half-spinor
representations and the vector representation exits only
for  $D_{4}$ $Spin(8)$, with Coxeter-Dynkin diagram
$${\rm o} \rule[2pt]{0.17in}{0.02in}
\shortstack{o \\ I \\ o}
\rule[2pt]{0.17in}{0.02in} {\rm o}$$

$D_{4}$ $Spin(8)$ has fundamental representations of
dimensions

$$8 \rule[2pt]{0.17in}{0.02in}
 \shortstack{8 \\ I \\ 28}
\rule[2pt]{0.17in}{0.02in} 8$$

For $D_{4}$ $Spin(8)$ the 8-dimensional vector representation
is isomorphic to each of the 8-dimensional half-spinor
representations.

This is the triality automorphism, a unique property of the
$D_{4}$ $Spin(8)$ Lie algebra.

Also, just as in the case of $D_{5}$, the exterior product
of the two half-spinor representations, $+8$ $\wedge$  $-8$,
has the same dimension, 28, as the representation to which
they are both connected.

\vspace{12pt}

The Weyl group of a Lie algebra is the finite reflection group of
of its root vector polytope.  The Weyl group of $D_{4}$ is of
order $2^{4-1} \times S_{4}$ = $8 \times 24$ = $192$.

The exceptional nature of $D_{4}$ is illustrated by its
root vector polytope.  Since $D_{4}$ has rank 4, the root
vector polytope is 4-dimensional.  It forms a 24-cell,
denoted by \{ 3,4,3 \} , that can be given quaternionic coordinates

$\pm 1, \pm i, \pm j, \pm k, and (\pm 1 \pm i \pm j \pm k)/2$.

These are the unit integral quaternions, and 4-dimensional
spacetime can be tiled with them, forming the $D_{4}$
lattice of integral quaternions \cite{COX}.

The $D_{4}$ lattice should be useful in forming a Feynman
checkerboard model in a 4-dimensional spacetime.

\vspace{12pt}

\subsection{$E_{6}$, $E_{7}$, and $E_{8}$}

\vspace{12pt}

For the $E_{6}$ Lie algebra, the Coxeter-Dynkin Diagram is:
$${\rm o} \rule[2pt]{0.17in}{0.02in} {\rm o}
\rule[2pt]{0.17in}{0.02in}
\shortstack{o \\ I \\ o}
\rule[2pt]{0.17in}{0.02in} {\rm o}
\rule[2pt]{0.17in}{0.02in} {\rm o}$$

Each node corresponds to one of the 6 fundamental representations
of $E_{6}$.  Their dimensions are:

$$27 \rule[2pt]{0.17in}{0.02in} 351
\rule[2pt]{0.17in}{0.02in}
\shortstack{78 \\ I\\ 2925}
\rule[2pt]{0.17in}{0.02in} 351
\rule[2pt]{0.17in}{0.02in} 27$$

The adjoint representation of $E_{6}$ is the $78$.

The two $351$s are each $27$ $\wedge$ $27$.

The $2925$ is  $27$ $\wedge$ $27$ $\wedge$ $27$.

Note that $78$ $\wedge$ $78$ = $3003$ = $2925 \oplus 78$.

\vspace{12pt}

The Weyl group of $E_{6}$ is the finite reflection group of a
6-dimensional (not regular) polytope with 72 vertices.

It is isomorphic to the group of automorphisms of the 27 lines
on the cubic surface in $\bf CP^{3}$.
Its order is $72 \times 6!$ = $51840$.

\vspace{12pt}

The 6-dimensional lattice formed by the $E_{6}$ root vectors
can be seen as the sublattice of the 8-dimensional $E_{8}$ lattice
that is the set of vectors in the $E_{8}$ lattice perpendicular to
any hexagonal $A_{2}$ sublattice.

The $E_{8}$ lattice is the 8-dimensional lattice of integral
octonions.  It can be formed from two $D_{8}$ lattices, much
as the 3-dimensional structure of diamond can be formed from
two $D_{3}$ lattices.

Each vertex of the $E_{8}$ lattice has 240 nearest neighbors.
They do not form a regular polytope in 8 real dimensions,
but they do form the regular Witting polytope in 4 complex
dimensions.

The Witting complex polytope has 40 diameters which
can represent the 40 root vectors of $D_{5}$ $Spin(10)$.

Each of the 40 hyperplanes orthogonal to one of the
40 diameters contains 72 vertices, which can represent the
72 root vectors of $E_{6}$.

The van Oss polygon of Witting polytope consists of the
vertices in a complex plane joining a complex edge to
the center.  It contains 24 vertices, which can represent the
24 root vectors of $D_{4}$ $Spin(8)$.

This material on polytopes and lattices is mostly taken
from Coxeter \cite{COX} and from Conway and Sloane \cite{CON}.

\vspace{12pt}

Unlike the $A_{n}$ and $D_{n}$ Lie algebras, which
can be represented by anti-hermitian complex matrices
or by anti-symmetric real matrices, the exceptional
Lie algebras $E_{6}$, $E_{7}$, and $E_{8}$ are based on
non-associative octonions and are not representable
in a straightforward way by associative matrices.

\vspace{12pt}

A rough, non-rigorous way to visualize
the Lie algebras $E_{6}$, $E_{7}$, and $E_{8}$ is:

\vspace{12pt}

$E_{6}$ = $Spin(10)$ $\oplus$  Spinor($Spin(10)$)
$\oplus$ $U(1)$,

where the dimensions are  78 = 45 + 32 + 1;

\vspace{12pt}
$E_{7}$ = $Spin(12)$ $\oplus$  Spinor($Spin(12)$)
$\oplus$ $SU(2)$,

where the dimensions are  133 = 66 + 64 + 3; and

\vspace{12pt}

$E_{8}$ = $Spin(16)$ $\oplus$  Half-Spinor($Spin(16)$),

where the dimensions are  248 = 120 + 128.

\vspace{12pt}

The fundamental representations of the exceptional Lie algebra
$E_{8}$ have been described by J. Frank Adams \cite{ADM}.
I understand that J. Frank Adams's book on exceptional Lie groups
may be forthcoming in the near future.

Two good treatments of the exceptional Lie algebras are
the Caltech preprint of Pierre Ramond \cite{PRM} and
the paper of A. Sudbery \cite{SUD}.

\vspace{12pt}

\section{Triality and Generalized Supersymmetry}

\vspace{12pt}

\subsection{$A_{n}$ $SU(n+1)$ does not work.}

\vspace{12pt}

Except for the low-dimensional cases of isomorphism with $D_{n}$:

$D_{2}$ $Spin(4)$ = $A_{1} \times A_{1}$ $SU(2) \times SU(2)$,

and

$D_{3}$ $Spin(6)$ = $A_{3}$ $SU(4)$,

the $A_{n}$ $SU(n+1)$ Lie algebras have only vector or multivector
fundamental representations, and do not have spinor representations.

Therefore, I do not think that the $A_{n}$ $SU(n+1)$ Lie algebras
are the fundamental mathematical structures on which to build
a theory of physics with both bivector gauge bosons and spinor
fermions.

The low-dimensional isomorphism cases are dealt with
systematically under the subsection on $D_{n}$ $Spin(2n)$.

\vspace{12pt}

\subsection{What about $D_{n}$ $Spin(2n)$?}

\vspace{12pt}

The $D_{n}$ $Spin(2n)$ Lie algebras all have both
bivector representations that can be used for gauge
bosons and two mirror image half-spinor representations
that can be used for fermion particles and fermion
antiparticles.

\vspace{12pt}

Is there a natural value of n for which the $D_{n}$ $Spin(2n)$
Lie algebra can be used to build a realistic model of physics?

\vspace{12pt}

\subsubsection{Conventional 1-1 supersymmetry, or
full spinor - bivector supersymmetry, does not work.}

\vspace{12pt}

Conventional 1-1 supersymmmetry between full spinor
fermions and bivector gauge bosons is the most obvious
criterion for selecting a natural value of n.

Since the spinor fermions are in the two half-spinor
representations of $D_{n}$ $Spin(2n)$, and each half-spinor
representation has dimension $2^{n-1}$, there are
$2^{n-1}$ fermion particles and $2^{n-1}$ fermion antiparticles.
By using the St\"{u}ckelberg-Feynman interpretation of
antiparticles as particles travelling backwards in time,
a supersymmetry between half-spinors and either
bivector gauge bosons or vectors whose exterior product
$vector$ $\wedge$ $vector$ = bivector gauge bosons
should be just as useful as a supersymmetry between full
spinors and either bivectors or vectors.

Since the gauge bosons are in the bivector representation
with dimension n(2n-1), there are n(2n-1) of them.

The following table shows the numbers of half-spinors,
of full spinors, and of bivector gauge bosons for the
$D_{n}$ $Spin(n)$ Lie algebras from n=1 to n=9:

\[
\begin{array}{|c|c|c|c|c|}
\hline
n & Group & Half-Spinors & Spinors & Gauge \: Bosons \\
\hline
1 & Spin(2) & 1 & 2 & 1 \\
\hline
2 & Spin(4) & 2 & 4 & 6 \\
\hline
3 & Spin(6) & 4 & 8 & 15 \\
\hline
4 & Spin(8) & 8 & 16 & 28 \\
\hline
5 & Spin(10) & 16 & 32 & 45 \\
\hline
6 & Spin(12) & 32 & 64 & 66 \\
\hline
7 & Spin(14) & 64 & 128 & 91 \\
\hline
8 & Spin(16) & 128 & 256 & 120 \\
\hline
9 & Spin(18) & 256 & 512 & 153 \\
\hline
\end{array}
\]

If n is larger than 9, it is clear that the dimension
$2^{n-1}$ of the half-spinor representations is much
larger than either of

2n, the dimension of the vector representation, or

n(2n-1), the dimension of the bivector adjoint representation.

\vspace{12pt}

There is no value of n for which the number of full spinors is
equal to the number of bivector gauge bosons.

Therefore, conventional 1-1 fermion-boson
supersymmmetry, or \\
full spinor - bivector supersymmetry, does not give a natural
value of n.

\vspace{12pt}

\subsubsection{Half-spinor - bivector supersymmetry does not work.}

\vspace{12pt}

Using the St\"{u}ckelberg-Feynman interpretation of
antiparticles as particles travelling backwards in time,
try to identify half-spinors 1-1 with bivector gauge bosons.

{}From the table, that works only for $D_{1}$ $Spin(2)$ = $U(1)$
acting on a 2-dimensional vector spacetime.

Therefore, Geisteswissenschaft indicates that a
$Spin(2)$  = $U(1)$ gauge field theory over
a 2-dimensional spacetime is natural, and indeed a
uniquely nice model can be built with $D_{1}$:

the 2-dimensional Dirac equation with its natural
Feynman checkerboard.

However, a $U(1)$ gauge field theory over a 2-dimensional
spacetime is obviously inadequate to describe
Naturwissenschaft.

The conformal field theory used in superstrings can be said to
be based on $D_{1}$, but it does not produce a single unique
theory that is consistent with Naturwissenschaft experiments.

Therefore conventional supersymmetry of fermions with
bivector gauge bosons has failed to satisfy the demands of
Naturwissenschaft experiments.

\vspace{12pt}

\subsubsection{Full spinor - vector supersymmetry almost works.}

\vspace{12pt}

Next, consider a generalized supersymmetry between
fermions and vectors, with the relationship to bivector
gauge bosons determined by bivector = $vector$ $\wedge$ $vector$.

A 1-1 symmetry between full spinors and vectors, according
to the table, works only for
$D_{2}$ $Spin(4)$ = $SU(2) \times SU(2)$
acting on a 4-dimensional vector spacetime.

Since $Spin(4)$ is the Lie algebra of the Lorenz group, and
Naturwissenschaft spacetime is 4-dimensional, the
1-1 full spinor - vector symmetry does give special relativity
and the 4-dimensional Dirac gamma matrices (full spinors)
for fermions.

Also, since $Spin(4)$ = $SU(2) \times SU(2)$ is reducible
to two copies of $SU(2)$, the 1-1 full spinor - vector symmetry
indicates that SU(2) gauge field theories over a 4-dimensional
base manifold spacetime should be interesting and useful,
as they are.

However, the $D_{2}$ $Spin(4)$ models are not elaborate
enough to explain by themselves the Naturwissenschaft
phenomena of $SU(3)$ color gauge forces and the 8 types
of fermion particles:  electron; e-neutrino; red, blue, and
green up quarks; and red, blue, and green down quarks.

\vspace{12pt}

Even though I don't think that $D_{2}$ models are big enough
to explain Naturwissenschaft, there are at least two lines
of work that have a chance at doing so successfully.

The earlier effort is that of Heisenberg and his
coworkers, including Durr and Saller, \cite{HSN}
to construct a unified field theory based on
$SU(2)$, an irreducible component of
$D_{2}$  $Spin(4)$ = $SU(2) \times SU(2)$.

The Heisenberg approach deals with the $SU(3)$ color force
and quarks by noting that the gluons and quarks are not
asymptotic states, but protons and pions are.  Therefore, protons
and pions should be represented by nonlinear states, such as
solitons, with the color force being represented as a higher-order
force generated by a fundamental urfield derived from $SU(2)$.

Whether or not the Heisenberg approach is successful in
all details, I think that it is correct in considering protons
and pions as soliton-type states, thus explaining the success
of phenomenological models such as bag models and
non-relativistic potential models using constituent quark
masses.

\vspace{12pt}

The later effort is that of Hestenes, Keller, et. al. \cite{HES}.
They notice that full spinors are defined as
$2^{n}$-dimensional minimal ideals in
$2^{2n}$-dimensional Clifford algebras, so that
full spinors can be defined as either left-ideals
or right-ideals.

In terms of $2^{n} \times 2^{n}$ matrices, the full
spinors can be chosen to be either \\
$2^{n} \times 1$ column vectors or $1 \times 2^{n}$
row vectors.

To use the whole Clifford algebra, both the left-ideal
column spinors and the right-ideal row spinors should
be used.

They then use the left-ideal column spinors to define
the spinor transformations of the vector spacetime,
and the right-ideal row spinors to define the fermion
spinor particles and antiparticles.

I think that their general approach is correct, but that the
4-dimensional full spinors of $D_{2}$ $Spin(4)$ are not
big enough for the 4-dimensional right-ideal row spinors
to account for all 8 types of fermion particles:
electron; e-neutrino; red, blue, and green up quarks; and
red, blue, and green down quarks.

\vspace{12pt}

\subsubsection{Half-spinor - vector supersymmetry does work.}

\vspace{12pt}
Using both the St\"{u}ckelberg-Feynman interpretation of
antiparticles as particles travelling backwards in time,
and a generalized supersymmetry between
fermions and vectors, with the relationship to bivector
gauge bosons determined by bivector = $vector$ $\wedge$ $vector$,
try to identify half-spinors 1-1 with
the vector representation.

{}From the table, that works only for in the unique
case with the property of triality:  $D_{4}$ $Spin(8)$
acting on an 8-dimensional vector spacetime.

\vspace{12pt}

To make a Naturwissenschaft physics model from
the 4 fundamental representations of $D_{4}$ $Spin(8)$,
it is natural to construct a Lagrangian gauge field model
with

the 8-dimensional vector representation as spacetime,

the 8-dimensional +half-spinor representation as the
fermion particles,

the 8-dimensional half-spinor representation as the
fermion anti-particles, and

the 28-dimensional bivector representation as the gauge
bosons.

\vspace{12pt}

The 8-dimensional vector spacetime can be reduced to
a 4-dimensional spacetime.

\vspace{12pt}

Prior to dimensional reduction, the generalized supersymmetry
relationship between the 28 gauge bosons and the 8-dimensional
spaces of fermion particles and antiparticles {\it might} give
an ultraviolet finite model.

In this connection, it is useful to note that, in an
8-dimensional spacetime, the dimension of each of the
28 gauge bosons in the Lagrangian is 1, and

the dimension of each of the 8 fermion particles is 7/2,
so that

the total dimension of the gauge bosons is equal to the
total dimension of the fermion particles, since
$28 \times 1 = 8 \times 7/2$.

\vspace{12pt}

After dimensional reduction of spacetime to 4 dimensions,

the fermions get a 3-generation structure and

the gauge bosons are decomposed by Weyl group
symmetries to produce

U(1) electromagnetism,

SU(3) color force,

SU(2) weak force plus a minimal Higgs field for
weak symmetry breaking, and

a Spin(5) = Sp(2) gauge field that can produce
gravity by the MacDowell-Mansouri mechanism.

\vspace{12pt}

\section{Conclusion:  $D_{4} - D_{5} - E_{6}$}

\vspace{12pt}

The program of half-spinor - vector generalized supersymmetry
has already been worked out in some detail \cite{SM}
(including hep-th/9302030 and hep-ph/9301210).

In particular, the structure of complex homogeneous
domains of hermitian symmetric spaces related to
the various gauge groups is used to calculate the
relative strengths of the forces, giving results
such as $\alpha_{E}$ = 1/137.03608 (see hep-th/9302030).

To get this structure, it is more natural to use the complex
and conformal structure \\
$E_{6}$ = $Spin(10)$ $\oplus$ Spinor($Spin(10)$)
$\oplus$ $U(1)$ = \\
=$Spin(8)$$\oplus$($\bf C$$\otimes$Vector($Spin(8)$))
$\oplus$$U(1)$
$\oplus$($\bf C$$\otimes$Spinor($Spin(8)$))$\oplus$$U(1)$

\vspace{12pt}

than it is to use the real structure \\
$F_{4}$ = $Spin(9)$ $\oplus$ Spinor($Spin(9)$) = \\
= $Spin(8)$ $\oplus$ Vector($Spin(8)$) $\oplus$
Spinor($Spin(8)$).

\vspace{12pt}

$E_{6}$ unites all 4 fundamental representations of
$D_{4}$ $Spin(8)$ using complex conformal structure, as:

$E_{6}$ is the conformal group whose Lorenz group
is $D_{5}$ $Spin(10)$, with the quotient space being
the rank 2 projective geometry of the spinor space of $D_{5}$
$Spin(10)$,  which spinor space in turn can be represented
as the complexification of the spinor space of $D_{4}$ $Spin(8)$;
and

$D_{5}$ $Spin(10)$ is the conformal group whose Lorenz
group is $D_{4}$ $Spin(8)$, with the quotient space being
the Lie sphere geometry of the complexification of the
vector space of $D_{4}$ $Spin(8)$.

\vspace{12pt}

$F_{4}$ unites all 4 fundamental representations of
$D_{4}$ $Spin(8)$ using real structure, as:

the quotient space $F_{4}$ / $Spin(9)$ is the
Cayley projective plane, which represents the
full spinor space of $Spin(9)$, which in turn can
represent the full spinor space of $D_{4}$ $Spin(8)$, and

the quotient space $Spin(9)$ / $Spin(8)$ is the
8-sphere $S^{8}$, which represents the vector
space of $D_{4}$ $Spin(8)$.

\vspace{12pt}

Roughly speaking, $F_{4}$ represents the 4 fundamental
representations of $D_{4}$ $Spin(8)$ in terms of the real
geometry of an ordinary sphere and a rank-1 projective space,
while $E_{6}$ represents them in terms of the complex
geometry of a Lie sphere and a rank-2 projective space.

\vspace{12pt}

The $E_{6}$ representation is the natural and correct one,
but I did not realize that until recently.

\vspace{12pt}

My earlier work used the $F_{4}$ representation because its
structure was simpler, and then I didn't know any better.
My $F_{4}$ model gave the same results as the $E_{6}$ model
because I added on the complex structure "ad hoc" rather than
realizing that it was the natural consequence of using
fundamental sets, quivers, and  $A-D-E$.

\vspace{12pt}

Therefore, the quantitative results of earlier papers
\cite{SM}, including but not limited to
$\alpha_{E}$ = 1/137.03608 and
a t-quark mass of 130 GeV, \\
come from a $D_{4}$ - $D_{5}$ - $E_{6}$ model of
Naturwissenschaft physics that is the natural
consequence of Geisteswissenschaft structures:

the empty set $\emptyset$ and the operation $\{ \}$;

quivers of sets and arrows;

the $A-D-E$ classification;

triality; and

half-spinor - vector generalized supersymmetry.

\end{document}